\documentclass[9pt,twocolumn,twoside]{osajnl}

\journal{ol} 

\setboolean{shortarticle}{true}

\newcommand{\micron}{$\mu\mathrm{m}\ $}

\newcommand{\redline}[1]{ {\color{black}#1}}

\title{Real-time ultrasound sensing with a mode-optimized photonic crystal slab}

\author[1,2,*]{Eric Y. Zhu}
\author[2]{Maria Charles-Herrera}
\author[1,2]{Cory Rewcastle}
\author[1,2]{Raanan Gad}
\author[1]{Li Qian}
\author[1,2,$\dagger$]{Ofer Levi}

\affil[1]{Department of Electrical and Computer Engineering, University of Toronto, 10 Kings College Road, Toronto, Ontario M5S 3G4, Canada}
\affil[2]{Institute of Biomedical Engineering, University of Toronto, 164 College Street, Toronto, Ontario M5S 3G9, Canada}

\affil[*]{eric.zhu@utoronto.ca}
\affil[$\dagger$]{ofer.levi@utoronto.ca}

\begin{abstract}
Integrated photonic sensors can provide large scale, flexible detection schemes.
Photonic crystal slabs (PCS) offer a miniaturized platform for wideband, sensitive ultrasound detection by exploiting the photoelastic effect in water. 
However, poor modal overlap with the sensing medium and non-negligible absorption loss of the aqueous medium have previously limited PCS sensor performance. In this study, we detail the development and optimization of a PCS-based acoustic sensor, by adding to it a low-loss high-index polymer cladding layer.
\redline{
Exploiting a mode-optimized TM-like optical resonance present in a PCS, with high bulk index sensitivity (>600 nm/RIU) and quality factor Q (>8000),  we demonstrate real-time 
ultrasound-sensing at a noise equivalent pressure (NEP) of 170 Pa (1.9 Pa/$\sqrt{\mathrm{Hz}}$).}
The PCS sensor is backside-coupled to optical fiber which, along with its intensity-based ultrasound-sensing architecture, will allow us to scale up easily to a 2D array.  
This work paves the way to a sensitive compact ultrasound detector for photoacoustic-based diagnostics and monitoring of tissue.
\end{abstract}

\setboolean{displaycopyright}{true}

\begin{document}
\maketitle

There is a need for robust, micron-scale ultrasound sensors that are also broadband and sensitive.  
Such sensors find uses in photoacoustic imaging, where intense laser light at specific wavelengths can be used to generate broadband acoustic waves in biological tissue; these acoustic signals,  when measured, give not only information about the gross anatomy of the tissue, but also allow for functional mapping such as blood oxygenation 
\cite{li2018photoacoustic} and tumor detection 
\cite{mehrmohammadi2013photoacoustic}.  
The availability  of a mass-manufacturable, broadband, and sensitive miniaturized acoustic sensor could also allow for miniaturized photoacoustic microendoscopy and fluorescence measurements in the same device \cite{MezilBOE2020}. 

The requirements above, unfortunately cannot be currently met by conventional piezoelectric transducers, which often have poor sensitivity at frequencies far from their resonant frequencies; they must also trade off sensitivity with size.  
While capacitive micromachined ultrasound transducers (CMUT) alleviate this issue somewhat, both piezoelectric and CMUT devices also suffer from high electromagnetic interference (EMI) and crosstalk \cite{zhou2003approach_piezo, zhou2007_CMUT}.  

In light of the shortcomings of conventional ultrasound sensors, recent work \cite{dong2017optical} has made all-optical ultrasound sensors a more attractive avenue of pursuit. 
In additional to their immunity to EMI, photonic ultrasound sensors have shown large measurement bandwidths and sensitivities (mPa/$\sqrt{\mathrm{Hz}}$) rivalling the best piezoelectric transducers but at much smaller sensor sizes \cite{guggenheim2017ultrasensitive,shnaiderman2020submicrometre, westerveld2021sensitiveNATPHOTON2021}.  
The types of optical sensors include fiber-Bragg grating-based \cite{rosenthal2014sensitive} and fiber-tip-based \cite{guggenheim2017ultrasensitive} devices, as well as planar integrated silicon-photonics circuits \cite{shnaiderman2020submicrometre, Hazan2018}.  
However, in many of these architectures, such as \cite{rosenthal2014sensitive, shnaiderman2020submicrometre, Hazan2018}, it may be difficult to create two-dimensional sensor arrays that are required for imaging.  
In our case, the use of photonic crystal slabs (PCS) as ultrasound sensors and compact, array-compatible  interrogation can allow for 2-D sensor array integration.

\begin{figure*}[t!]
\centering
\includegraphics[width=17cm]{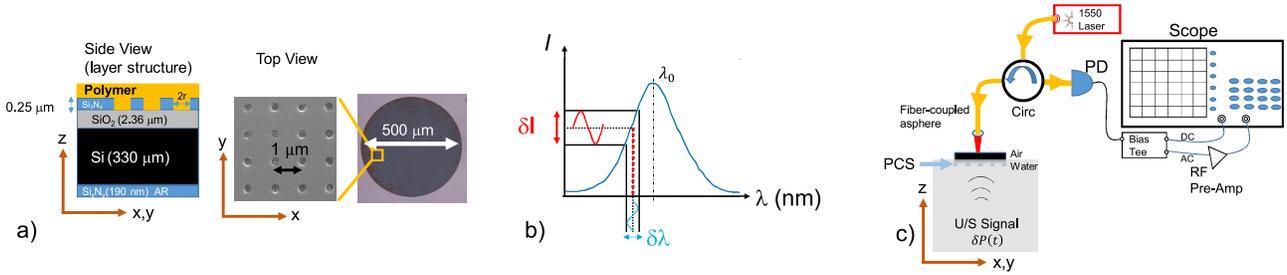}
\caption{
\label{fig:setup}
a) Layer structure of the PCS sensor.  
b) The optical resonance.  The index-sensing nature of the PCS means that the resonance shifts in wavelength $\delta\lambda$ when the index of the overlayer $n_M$ is changed.  This causes a commensurate shift $\delta{I}$ in the reflectance of the PCS.  
c) Experimental setup.  The PCS is inverted and immersed in water; the backside is optically interrogated by a 1.5-\micron tunable laser using a fiber-coupled asphere, while ultrasound (U/S) signals are incident on its topside.  
Devices: circulator (Circ), photodiode (PD), (oscillo)scope.  
\vspace{-10pt}
}
\end{figure*}

In previous work \cite{zhu2019refractive}, we demonstrated the ability to sense broadband ultrasound signals in water using PCS sensors (Fig. \ref{fig:setup}a).  
The (TE-like) optical resonance in the PCS shifts in wavelength (Fig. \ref{fig:setup}b) 
due to the change in refractive index $\delta{n}$ of the sensing medium induced by an ultrasonic  (pressure) signal $\delta{P}$, and this allowed us to map that signal onto the intensity $\delta{I}$ of an interrogating optical beam.  
The mechanism can be succinctly summarized with the following equation:

\begin{equation}
	\delta{I} = (1+r) \times  \delta{P} \times \frac{dn_M}{dP} \times \frac{d\lambda }{dn} \times \frac{dI}{d\lambda}, 
	\label{eqn:OriginalMechanism}
\end{equation}
\noindent
where $r\sim 1$ is the acoustic Fresnel coefficient at the water-PCS interface, $\frac{dn_M}{dP}$ is the photoelastic coefficient of the overlayer (in this case, water) surrounding the PCS, $S = \frac{d\lambda }{dn}$ is the bulk index sensitivity of the PCS, and $\frac{dI}{d\lambda}$ is a quantity that depends on the sharpness of the resonance (linewidth, $\gamma$) and the amount of light coupled out of the resonance $I_0$ $\left(\frac{dI}{d\lambda}\sim \frac{I_0}{\gamma}\right)$.

In this work, we achieve real-time measurement of ultrasound signals by interrogating a TM-like guided mode resonance \cite{johnson1999guided} in the PCS, while still exploiting the same mechanism given in Eqn \ref{eqn:OriginalMechanism}.
\redline{
We observe 
a noise equivalent pressure (NEP) 
of 1.9 Pa/$\sqrt{\mathrm{Hz}}$, 3.8-fold better than our previous work \cite{zhu2019refractive}.
}

TM-like guided mode resonances in PCS are ideal for index-sensing due to their higher quality factors $Q$ 
($\equiv \frac{\lambda_0}{\gamma}$, in excess of $10^4$) and higher $S$ values compared to their TE-like  counterparts \cite{Checkerboard_Nicolaou:13}.  
However, the absorption loss of water-based analyte solutions has always limited the measurable $Q$ values of these resonances.  
Additionally, TM-like modes are much more difficult to couple light into at normal incidence than TE-like modes in PCS.  
Here, we overcome these challenges by (1) applying a low optical absorption  polymer overlayer onto the PCS; this allows the TM resonance to maintain its high $Q$ (8000), while the polymer's presence increases the mode-field energy in the sensing medium and results in a higher $S$ (670 nm/RIU).
The two polymers used (PMMA, BCB) have material absorption  $\alpha$ (Table  \ref{Tab:Results})  much smaller than that of water \cite{PMMA_Loss_kaino2014optical, BCB_Loss_Gassenq:12,Water_Loss_Palmer:74}.  
We also (2) improve  coupling into the TM-like optical resonance by more than 10-fold using a miniature fiber-coupled asphere (Fig. \ref{fig:setup}c).

The layer structure of the PCS  is shown in Fig. \ref{fig:setup}a.  It consists of a high-index guiding layer (stoichiometric silicon nitride, Si$_3$N$_4$, $n$= 2.0), sandwiched between a lower-index bottom cladding of fused silica (SiO$_2$, $n = 1.45$) and a top cladding layer $n_M$ ($1<n_M<1.6$) such as air, water, or some other material.  
In this guiding  layer, a periodic square lattice of circular nanoholes (Fig. \ref{fig:setup}a), patterned with electron-beam lithography and etching, allows for the presence of optical Bloch modes.   
It is these Bloch modes which give rise to the optical resonances present in the PCS (Fig. \ref{fig:setup}b) \cite{joannopoulos2008molding}.  
The lattice constant ($a \sim 1.0\ \mu\mathrm{{m}}$) of the periodic structure is chosen so that these guided mode resonances occur near 1550 nm.  
\redline{A thin layer of PECVD silicon nitride (190 nm) is deposited onto the backside of the substrate to act as an antireflection (AR) coating \cite{el2010sensitivity} to enable backside coupling.  }

A polymer overlayer, approximately 1.8 $ \ \mu\mathrm{m}$ thick, is applied to the PCS through spin-coating and soft-baking.  
Much more of the mode field energy lies in the overlayer of the PCS for the TM-like mode compared to the TE-like mode.  
{
The bulk index sensitivity $S$ value of each resonance increases with the fraction of field energy that lies in the overlayer (with index $n_M$) \cite{el2010sensitivity, joannopoulos2008molding}:  

\begin{equation}
S\equiv \frac{d\lambda}{d n_M} = \frac{\lambda_0}{n_M} \times \frac{\int_{\mathrm{over}}{d^3\vec{x} \ n^2_M |\vec{E}(\vec{x})|^2}}
{ \int d^3\vec{x}\ n^2(\vec{x}) |\vec{E}(\vec{x})|^2}.  
\label{eqn:BulkSens}
\end{equation}

\noindent
While in previous work \cite{zhu2019refractive}, we utilized only TE-like resonances, whose mode fields had limited penetration into the overlayer $n_M$ and correspondingly lower $S$ and acoustic sensitivities, in this work we use a TM-like resonance.  
The higher index $n_M$ overlayers also drag out the mode-field energy into the overlayer, which creates in turn a higher $S$ for the optical resonance (Eqn \ref{eqn:BulkSens}). 
We see this pictorially in Figures \ref{fig:xsection}a--c; a cross-section of the mode-field energy in a single unit cell is shown when different $n_M$ are used.  
Figure 2d shows the overall trend for peak wavelength and $S$ as the sensing index
$n_M$ is increased from 1.3 to 1.6.  
The higher indices of $n_M = 1.48$ and $n_M = 1.54$ correspond to different polymers (PMMA and BCB, respectively) that are applied to the PCS.  
The values of $S$ achieved with these two polymers (Table \ref{Tab:Results}) 
are within striking distance of the upper bound of $S$ ($\leq \lambda/n_M\sim 1000$ nm/RIU) 
\cite{UltimateS_Yu:11}.

\begin{figure}[t!]
	\centering
	\includegraphics[width=8.5cm]{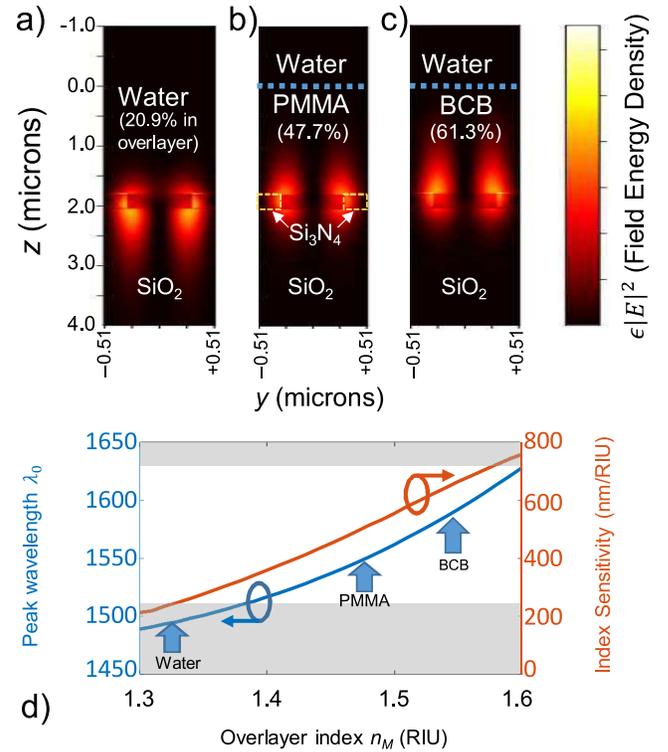}
	\caption{
		\label{fig:xsection}
		Cross-sections of the TM mode-field energy when various overlayers of thickness $\sim 2$ \micron are used: a) water, b) PMMA, and c) BCB.  
		d)  The peak wavelength (blue) and $S$ are plotted as a function of the overlayer 
		index $n_M$.
		The indices of refraction for the various overlayers we use are denoted with arrows.   
		As $n_M$ increases, so does $\lambda_0$ and $S$.   
		\vspace{-10pt}
	}
\end{figure}

Increasing the $S$ of an optical resonance, however, is not sufficient to improve the acoustic sensitivity of our PCS sensor, as the other properties of the overlayer and sensor may also change (Eqn \ref{eqn:OriginalMechanism}), including the acoustic Fresnel coefficient $r$ and photoelastic coefficient $\frac{dn_M}{dP}$.  Table 1 lists these properties for water and the two polymers \cite{waxler1963effect,WaxlerPMMA,HossainBCB}.  

Taking the direction of the applied stress to be in the $z$ direction (Fig. \ref{fig:setup}c), 
$\frac{dn_{z}}{dP}$ ($\frac{dn_{x,y}}{dP}$) is the photoelastic coefficient parallel (perpendicular) to $z$.  These two values do not, in general, equal each other, as is the case for the two polymers (Table \ref{Tab:Results}).  
Their values also happen to be smaller than that of water, which has identical (isotropic) coefficients.  
Since we are only using TM modes in the PCS for sensing, where most of the mode field lies in the $z-$direction, we can simply use $\frac{dn_{M,z}}{dP}$in Table \ref{Tab:Results}.  

The acoustic Fresnel coefficient at the polymer-PCS interface $r$ (Eqn \ref{eqn:OriginalMechanism}) is determined by performing a 1-D fluid-dynamic FDTD simulation in k-Wave \cite{treeby2010kWave}.  We find it remains $\sim 0.86$ for both PMMA and BCB; {this is  expected because 
(1) their acoustic impedances are similar to that of water, and (2)
their layer thicknesses ($t\sim 2\ \mu{\mathrm{m}}$) are much smaller than the acoustic wavelength $\lambda_a$ --- for $f<$ 100 MHz, the acoustic wavelength $\lambda_a>$ 15 $\mu\mathrm{m}$ --- 
meaning that the ultrasound essentially sees the overlayer as
part of the water medium.}

We then experimentally assess the acoustic-sensitizing properties of the two different polymers
on the TM resonance in our PCS.  
We note that it is not possible to assess the ultrasonic sensing capabilities of the TM resonance when the PCS is directly immersed in water; this is because the center wavelength of the resonance (Fig. \ref{fig:xsection}d, grey region) lies outside the tunable range of our interrogating laser.  
The PCS sensor is immersed in a water tank with the polymer coating and PCS-patterned layers lying on the water surface (Fig. \ref{fig:setup}c).  
Underneath the PCS and completely immersed in water is an ultrasound transducer.
The back surface of the PCS is above the water line and interrogated with a 1.5-$\mu{m}$ tunable laser (Keysight 81960A).  
A miniature fiber-coupled aspheric collimator ($f = 4.5$ mm asphere)  takes the laser light and illuminates the PCS with a 900-\micron diameter spot.  
{
The backreflected beam from the PCS is then back-coupled into the fiber, and a circulator (Circ) directs the light to a fast photodiode (PD).  Right before the PD, a polarization controller and fiber-coupled polarizer (not shown in Fig.  \ref{fig:setup}c) is set to perform a cross-polarized measurement to remove Fabry-Perot fringes from the measurement of the resonance.
}

\begin{figure}[t!]
	\centering
	\includegraphics[width=8cm]{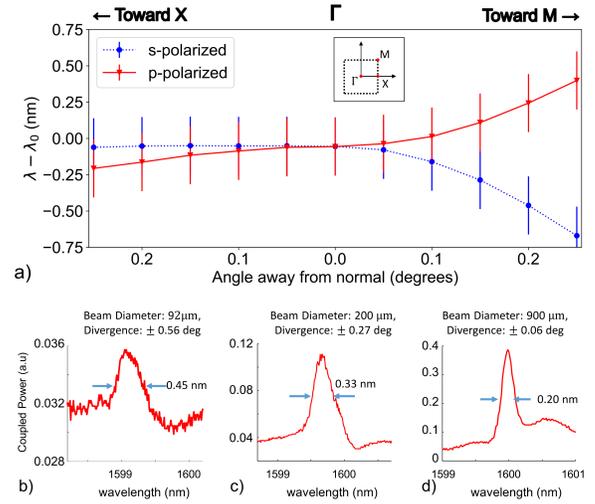}
	\caption{
a) The band structure of the TM-like guided mode resonance (GMR).  The mode consists of two bands 
 that intersect at normal incidence (the $\Gamma$ point).  	
The bands experience large angular dispersion; 
they diverge quickly as the angle of incidence deviates
from normal incidence. 
The vertical bars represent the expected linewidth when the device is interrogated with a plane wave with the given angle of incidence and polarization.
Inset shows first Brillouin zone of the square lattice, with  critical points X and M. 
b)-d)  Reducing the beam divergence of the interrogating beam increases both the amount of power coupled out of the GMR as well as the measured quality factor $Q$ of the resonance.
\redline{In all three cases, the PCS is covered in BCB, and the same power is used for the incident interrogating beam.}
Linewidths are extracted by fitting the resonances to a Fano peak.  
\label{fig:Bands_Diverge}
\vspace{-10pt}
}
\end{figure}

Due to the large angular dispersion of the band structure supporting the TM-like optical resonance (Fig. \ref{fig:Bands_Diverge}a), 
smaller interrogating beam divergences yield narrower measured linewidths.
\redline{
In Fig. \ref{fig:Bands_Diverge}b--\ref{fig:Bands_Diverge}d, we show for
a BCB-covered sensor that decreasing the beam divergence 
of the interrogating beam (Fig. \ref{fig:setup}c) 
allows us to couple almost 100 times more light out of the
resonance (Fig. \ref{fig:Bands_Diverge}d) compared to our previous setup  \cite{zhu2019refractive} (Fig. \ref{fig:Bands_Diverge}b);
it also increases the quality factor $Q$ of the measured resonance.  This is because
the interrogated resonance is less `spread out' over both k-space and frequency.}
The higher optical power coupled out of the resonance allows us to operate at well above the dark current level of our photodetector, significantly reducing detector-based noise.
A TE-like resonance in the same slab, on the other hand, accomodates much larger divergence beams for interrogation due to its much larger linewidth ($\sim$ 10 nm).  The larger $Q$ and higher $S$ of the TM-like resonance comes at the cost of challenging coupling.  


The acoustic sensitivity of the polymer-overlayer PCS is then tested.  
Gaussian-shaped ultrasound pulses (center frequency 10 MHz, 1 $\mu{s}$ FWHM pulsewidth, amplitude varying from 0 to 80 kPa) are normally incident on the PCS.  The signal at the photodiode is then collected with an oscilloscope and averaged over 64 pulses; a bias-tee is used to separate the low-frequency (DC--500 kHz) and recorded high-frequency (> 500 kHz) components.  A Gaussian fit is used to extract the amplitude of the signal $\delta I$ (in V), for the corresponding applied pressure ($\delta P$).  

\redline{
Similar to previous PCS sensors supported by 
TE-like resonances  \cite{zhu2019refractive,ZhuCLEO2019},  we observe that our current polymer-covered TM-resonance-based sensor also has a broadband response to ultrasonic signals.   A flat response is measured from 3--40 MHz, with the range limited only by our measurement apparatus.
}

In order to properly compare the effect of one polymer to another, we must account for variations in light coupling.  
By dividing both sides of Eqn (1) by $dI/d\lambda$ (the reflectance slope of the optical resonance), we can obtain the wavelength shift $\delta\lambda$ of the device as a function of the applied pressure $\delta P$.
%
%
\noindent
Figure \ref{fig:Sig_vs_Pressure}a plots the wavelength shifts for both BCB and PMMA.  
The wavelength shifts remain linear over the range of applied pressures.  
We also observe that the BCB-covered sensor is more than three times as sensitive ($d\lambda / dP = 0.11$ nm/MPa)  as the PMMA-covered sensor 
($d\lambda / dP = 0.034$ nm/MPa).  
These values can be compared to Eqn \ref{eqn:OriginalMechanism} by using the properties of the PCS and the known photoelastic coefficients $\frac{dn_M}{dP}$ of the polymers \cite{WaxlerPMMA,HossainBCB}.  The results are tabulated in Table \ref{Tab:Results}, and we find that the expected theoretical values of $d\lambda / dP$ are within 30\% of our experimentally-obtained values. 

We can also measure the minimum detectable pressure of our polymer-overlayer devices.  Only the data for the BCB-overlayer device is shown because it is more sensitive.  By reducing the applied pressure $\delta{P}$ 
gradually (Fig. \ref{fig:Sig_vs_Pressure}b), we can find where the slope of the signal intersects the measurement noise floor; this point, the NEP, happens to be 0.17 kPa.  The normalized NEP, which accounts for averaging (with a measurement time of 64$\times1$ $\mu{\mathrm{s}}$), is 1.9 Pa/$\sqrt{\mathrm{Hz}}$.  This is an improvement of more than 3.8-fold over our previously-reported sensor \cite{zhu2019refractive}.  

\begin{figure}
	\centering
	\includegraphics[width=8.5cm]{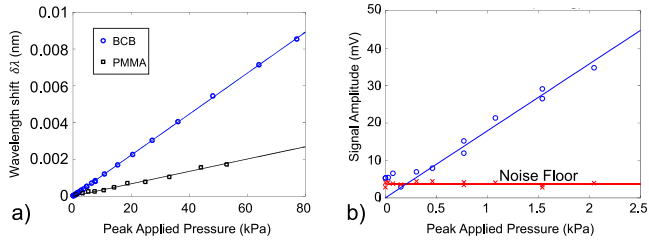}
	\caption{
	\label{fig:Sig_vs_Pressure}
	a) The wavelength shift vs applied acoustic pressure is shown for the PMMA- and BCB-covered PCS.  
	The  signal produced by the BCB sensor is more than three-fold stronger than the PMMA-overlayer device, and the $\frac{d\lambda}{dP}$ for each is within 30\% of the theoretical value (Table 1).  	
	b) Noise floor measurement for the BCB-covered sensor, indicating an NEP of 0.17 kPa.  
	\vspace{-10pt}
	}
\end{figure}


Previous work on silicon-on-insulator (SOI) waveguide-based ultrasound sensors (\cite{kumar2019enhanced}) also utilized polymer overlayers in place of a silica overcladding.  
That work exploited the increased photoelastic coefficient of the polymer (compared to silica) to obtain an improvement in the acoustic sensitivity, also with a TM mode in an Si slab waveguide.  
But due to the lack of an optical resonance, the sensing architecture in \cite{kumar2019enhanced} required the construction of a stabilized interferometer and phase-based measurements. In this work, we can simply utilize intensity-based measurements due to the presence of an optical resonance.

Our sensor interrogation topology has several benefits 
over traditional photonic-integrated circuit (PIC)-based sensors that require either expensive packaging (for grating couplers and edge coupling) or expensive stages (for butt coupling).  We can interrogate our sensor from the backside with a free-space beam, or simply backside-couple a small fiber-coupled lens, 
as we have done here (Fig. \ref{fig:setup}c); 
\redline{this latter configuration can be further miniaturized with a microlens array, with pitches as small as 200 \micron, to allow for the realization of a 2D PCS sensor array.  
}

In summary, we have used a TM-like resonance in our PCS device to make real-time measurements of ultrasound signals.
{
This was facilitated by applying a high-index low-loss polymer overlayer onto the PCS, whose high $Q$ optical resonance was maintained while its $S$ was increased (Eqn. \ref{eqn:OriginalMechanism}).
}
Additionally, the small beam-divergence of the interrogating lightwave increased the amount of light coupled out of the sensor.  
These approaches allowed us to improve the sensitivity of our photonic-crystal slab-based ultrasound sensors by almost four-fold over our previous work, down to an NEP of 1.9 $\mathrm{Pa}/\sqrt{\mathrm{Hz}}$.  
We expect further improvements in sensitivity (down to  $100\ \mathrm{mPa}/\sqrt{\mathrm{Hz}}$ levels) by further optimizing the $Q$ of the resonances.  
This intensity-based ultrasound-sensing architecture will allow us to scale up easily to a 2D array, and paves the way to a compact sensor for photoacoustic-based diagnostics and monitoring of tissue.  

{\noindent\bf Acknowledgments.}  We acknowledge support from NSERC, the Toronto Nanofabrication Centre (TNFC), 
and thank Prof Amir Rosenthal for lending us the BCB.  

{\noindent\bf Disclosures.}  The authors declare no conflicts of interest.

{\noindent\bf Data availability.} Data presented in this paper may be obtained from the authors upon reasonable request.

\begin{table}[h!]
\centering
\begin{tabular}{  c|c|c|c}
Material & Water & PMMA & BCB \\ \hline
%
%

%
Thickness (\micron) 		&   N/A  	&      1.74    	&  1.85 \\ 
\hline
%
$r$ (Acoustic)   & 0.86 & 0.86 &  0.86\\ \hline

$n_M$ ($\lambda = 1550$ nm)   						& 1.32 	& 1.48 	& 1.54  \\  
\hline
{$\alpha$ (dB/cm) { ($\lambda = 1550$ nm)  } } &   100  &   0.04 &   3.0  \\  
\hline
$dn_{x,y}/{dP}$ 	($\times{10^{-6}}$ RIU/MPa)		& 		$138$ 	&	$41$		&		$31$	       \\ 
${dn_{z}}/{dP}$   ($\times{10^{-6}}$ RIU/MPa)    			& 		$138$ 	&	$43$		&	    $99$	       \\ 
\hline
$S$ (nm/RIU) 	& 233 & 510	&   670\\ \hline
$\left({d\lambda}/{dP}\right)_{Calc}$  (nm/MPa) 	& 0.060   &  0.041	&   0.12\\  
$\left({d\lambda}/{dP}\right)_{Meas}$   (nm/MPa) 	& N/A      &   0.034	 &  0.11\\  
\hline
\end{tabular}
\centering
\caption{Experimental Conditions and Results with Different Overlayers\label{Tab:Results}
}
\end{table}

%
%

\end{document}